# IMPACTS OF REGIONAL TRADE AGREEMENTS (RTAS) ON FOOD SECURITY: A CASE OF ASEAN FREE TRADE AGREEMENT

**HERATH H.M.S.P.**[*]; **CAO LIANG**[**]; **CHEN YONGBING**[***]

[*]PHD SCHOLAR,
SCHOOL OF BUSİNESS ADMİNİSTRATION,
ZHONGNAN UNIVERSITY OF ECONOMICS AND LAW, WUHAN, P.R. CHINA.

[**]PROFESSOR,
SCHOOL OF BUSINESS ADMINISTRATION,
ZHONGNAN UNIVERSITY OF ECONOMICS AND LAW, WUHAN, P.R. CHINA.

[***]ASSOCIATE PROFESSOR,
SCHOOL OF BUSINESS ADMINISTRATION,
ZHONGNAN UNIVERSITY OF ECONOMICS AND LAW, WUHAN, P.R. CHINA.

**ABSTRACT**
Discriminatory trade liberalization policies are becoming more popular among world economies. Countries are motivated to enter for regional trade agreements to capture faster economic growth for alleviating poverty. In developing economies like most of the member countries of the Association of South East Asian Nations (ASEAN), a sizeable portion of people are suffering from poverty by exposing them to food insecurity. Low level of income and low productivity of agricultural sector have augmented the severity of food insecurity of those people. Discriminatory trade liberalization policies are expected to reduce poverty and strengthen the food security. The objective of this paper is to examine the effect of ASEAN Free Trade Agreement (AFTA) on food security of its member countries. The multiple regression analysis in panel data was employed to disentangle the impacts of trade liberalization on food security with use of regional trade agreement dummy variable. The finding of the study supports that AFTA has influenced positively on food security of its member nations. After the formation of AFTA, the level of per-capita daily dietary energy supply of the member countries has been increased moderately over time.

## 1. INTRODUCTION

With a motto of one vision, one identity and one community, the association of Southeast Asian Nations (ASEAN) was born on 8[th] August 1967 with five founding members, namely Indonesia, Malaysia, Philippines, Singapore and Thailand. Later another five members of the region namely Brunei Darussalam, Viet Nam, Lao PDR, Myanmar and Cambodia were joined by further strengthening the ASEAN association. Brunei Darussalam joined to the association on 7[th] January 1984, Viet Nam on 28[th] July 1995, Lao PDR and Myanmar on 23 July 1997 and Cambodia on 30[th] April 1999 making up ten member countries of ASEAN. Among several socioeconomic, cultural, political and security aims and purposes, major economic objectives of the ASEAN were to accelerate economic growth, collaborate more effectively for greater utilization of agricultural and manufacturing industries of member states and expand their trade.

With the aim of discriminatively reducing or eliminating tariff and other trade barriers, six members of ASEAN, Brunei Darussalam, Indonesia, Malaysia, Philippines, Singapore and Thailand, singed the ASEAN Free Trade Agreement (AFTA) in 1992 with entering into force in 1993. The new ASEAN member-countries entered into free trade agreement in 1995 by Vietnam, in 1997 by Laos and Myanmar and Cambodia in 1999 and they were given longer time period to reduce their tariff rates. The AFTA aimed to reduce and eliminate tariffs through the Common Effective Preferential Tariff (CEPT) plan with an aim of achieving zero tariffs for all products by 2010 for the six founding member states and 2015 for the CLMV which is the acronym for Cambodia, Laos, Myanmar and Viet Nam. After establishment of the AFTA, main objectives of member nations were to increase the competitiveness in the region and world markets by eliminating intra-ASEAN tariffs and non-tariff trade barriers (NTBs) and attract more foreign direct investments (FDIs) into the region.

Table 1: Contribution of Agriculture in ASEAN Nations – 2011

| Country | Agriculture Value Added (% of GDP) | Employment in Agriculture (% of Total Employment) |
|---|---|---|
| Brunei Darussalam | 0.64 | … |
| Cambodia | 36.68 | 72.2[c] |
| Indonesia | 14.72 | 38.2[a] |
| Lao PDR | 30.8 | 85.4[g] |
| Malaysia | 11.87 | 13.5[b] |
| Myanmar | 48.35[e] | 62.7[f] |
| Philippines | 12.79 | 35.2[b] |
| Singapore | 0.03 | 1.1[b] |
| Thailand | 12.36 | 41.5[b] |
| Vietnam | 22.02 | 51.7[d] |

a- 2010 b- 2009, c- 2008, d- 2006, e- 2000, f – 1998, g-1995
Source: World Development Indicators – United Nations

Majority member nations of ASEAN are recognized as developing countries. Agricultural sector of these member nations is still the most dominant sector and contributes a larger portion to national income and employment (Table 1). Majority people of these countries have been employed in farming and agricultural related works. As a result, the living standard of these people greatly depends on the development of agricultural sector. Poverty and hunger which are the cause and consequence of food insecurity is linked with agricultural sector of those economies.

In developing economies like most of the ASEAN members, a sizeable portion of people are suffering from severe poverty by exposing them to food insecurity. Low level of income and low productivity of agricultural sector have augmented the severity of food insecurity of people. As a result of that, policymakers of these developing countries are seeking means to get rid from the issue by making domestic and external reforms to solve the issue. While domestic reforms targeting on macroeconomic and agricultural sector reforms, trade reforms direct towards import and export policies. With increased interest on trade reforms in rapid globalization, policymakers of developing economies pay attention on trade liberalization to enhance the food security of their economies. Especially, international organizations during the last two decades are exploring links between trade liberalization and food security of world economies (FAO, 2003).

International economists expect that discriminatory and broader trade liberalization generate positive impacts on the agricultural sector of countries. As the agricultural sector is the dominant sector in most developing economies, any positive impact of RTAs is important to develop or enhance the economic performance of that sector. Trade policy reforms can affect prices of agricultural commodities. Moreover, production and trade volume of agricultural commodities are expected to be affected with trade reforms. These effects can impact on poverty and hunger and ultimately influence food security of trading members.

Table 2: Economic Indicators of ASEAN Members – 2011

| Country | Surface Area (in thousand sq. km) | Population (Million) | GDP Per Capita (US$) | Average Economic Growth (2000-2011) |
|---|---|---|---|---|
| Brunei | 5.77 | 0.41 | 40301 | -0.51 |
| Cambodia | 181.04 | 14.31 | 897 | 6.70 |
| Indonesia | 1904.57 | 242.33 | 3495 | 3.97 |
| Lao PDR | 236.8 | 6.29 | 1320 | 5.35 |
| Malaysia | 330.8 | 28.86 | 9977 | 2.95 |
| Myanmar | 676.59 | 48.34 | .. | .. |
| Philippines | 300 | 94.85 | 2370 | 2.77 |
| Singapore | 0.71 | 5.18 | 46241 | 3.69 |
| Thailand | 513.12 | 69.52 | 4972 | 3.44 |
| Viet Nam | 331.05 | 87.84 | 1407 | 6.00 |

Source: World Development Indicators – World Bank

Policymakers in developing countries face a policy dilemma. Trade liberalization polices have produced mixed bag of results on the economic performance of developing economies. Researchers have attempted to answer the question of how RTAs influence on economic performance of members and non members of trade blocs. Though many researches have been carried out on RTAs, there is a deficiency of literature linking regional trade agreements on agricultural trade and food security. Consequently, this study attempts to fill this gap by finding empirical evidence in relation to the impact of RTAs on agricultural trade and food security. The primary objective of the study is to disentangle the impacts of AFTA on food security of its member countries.

*Literature Review*

Though there is an increasing awareness on impact of trade liberalization on food security relatively there is a deficiency of research on this field. According to the World Food Summit held in 1996, food security is defined as "the existence of having all people at all times for sufficient, safe, nutritious food to maintain a healthy and active life" (World Health Organization, 2012). This definition explains physical and economic access to foods for all times. Further, the researchers have pointed out three important pillars in which food security is built on. *Food availability*, *food access* and *food use* are such pillars in ensuring food security of the people of any country. Food availability refers to persistent availability of sufficient quantities of foods for people. Food access explains that people must have enough resources to obtain nutritious diet in their daily life. These first two pillars elucidate physical and economic access to food needs of people. Food availability is determined mainly by domestic production, importation and food aid. However, food security is not enhanced only with fulfilling the food availability of an economy unless the food accessibility of people is satisfied. If people don't have adequate purchasing power even when the food availability is fulfilled it can cause to food insecurity. Prominent economist professor Amartya Sen has clearly exemplified in the Bengal famine of 1943. In this scenario, he explains that in Bengal famine even food was available people were starving because of the lack of purchasing power of people. As a result purchasing power is the major determinant of food accessibility. Consequently, employment level, economic growth, inflation, public distribution systems can influence on purchasing power of people (Darshini, 2012). The third pillar of food security which is the food use refers to People's adequate knowledge on appropriate use of basic nutrition and care and adequate water and sanitation facilities of food use under food security (FAO, 2003).

In Preparation of Comprehensive National Food Security Programme, Food and Agriculture organization define food security as the physical and economic access to adequate food for all household members, without undue risk of losing the access for food. The definition of World Bank on food security refers to access by all people at all times to enough food for active and healthy life (World Bank, 1986). According to World Bank, food availability and accessibility are the major elements of food security. Consequently, it refers food insecurity as the lack of access to enough food. Chronic food security and transitory food security are the main two forms of food insecurity. While chronic food insecurity is refereed as a continuously and inadequate diet caused by the inability to acquire food transitory food insecurity is defined as a temporary decline in household's access to enough food (World Bank, 1986). In *Post –Colonial State and food security in Southern Africa*, Mkandawire and Maltosa (1993) defined food security in most simple terms as the absence of hunger and malnutrition. By incorporating fundamental concepts of various definitions of food security, USAID define the food security as when all people at all times have both physical and economic access to sufficient food to meet their dietary needs for a productive and healthy life (USAID, 1992).

Bezuneh and Yiheyis (2009) have examined the food security in national level with special reference to the trade liberalization. The researchers have conceptualized food security as food availability for human consumption at the national level both from domestic production, commercial imports and food aid. Bezuneh and Yiheyis have made efforts to empirically investigate the effect of trade liberalization on food security. In their study, researchers have employed multiple-regression model to capture impacts of trade liberalization on food security.

The dependent variable of the regression model is referred as the food availability. Food availability is represented by per capita daily dietary energy supply and is derived from food balance sheets using country-level data on domestically produced and imported foods including food aid available for human consumption minus nonfood use. Researchers have estimated impact of trade liberalization on food security with the use of set of control variables. The set of control variables employed in their econometric model are per capita real GDP (RGDPPC), irrigated land as a percentage of crop land (IRG), the price of imported foods (MFPRICE), foreign reserves in months of imports (RESVM) and political instability (POL). In assessing impact of trade liberalization, researchers have employed trade liberalization dummy variables with three lags to capture time effect of trade liberalization.

Bezuneh and Yiheyis hypothesized a positive association between food security and trade liberalization with favorable impact of increased supply of food and decreased food prices. It is hypothesized a positive relationship between food security and real gross domestic product. Positive association between real GDP and food security is expected because of the favorable impact of increased income on food expenditure. Researchers hypothesized a positive association of food security and the percentage of irrigated crop land since increased of irrigated land directly contribute to enhance the domestic supply thereby increasing food availability. Also, amount of foreign reserves are assumed to be favorably improve the food security level through enhancing the ability of food importation of countries. However, researchers hypothesize negative associations between food imports prices and political instability with food security. Researchers explain that political instability creates unfavorable condition on food security through its impact on food supply from domestic production.

## 2. METHODOLOGY
*Data Description*
In identifying the impact of RTAs on food security, panel data is collected from 1980 to 2009 period. Data for real per capita GDP, agricultural land, imported food price, foreign reserves is collected from World Bank Development Indicators of World Bank data base. Data for per capita daily dietary energy supply is collected from Food and Agriculture Organization of the United Nations Statistical Database (FAOSTAT). The study use the Polity2 score as a measure for political instability which has been published in the Polity IV dataset project by Monty G. Marshall of the University of Maryland, College Park and Keith Jaggers of Colorado State University in 2002. Polity2 scores ranges from -10 (strongly autocratic) to +10 (strongly democratic).

*Methods of Data Analysis*
This study employs panel data regression technique in analyzing impacts of RTA on food security. In achieving major objective of the study dependent variable, Per Capita Dietary Energy Supply, is adopted to measure the food availability which is an approximate to food security. The multiple regression analysis in panel data is employed to disentangle the impacts of discriminatory trade liberalization on food security. A set of independent variables which influence on the dependent variable is included in the panel regression model. Impacts of ASEAN free trade agreement on food security is captured by adopting regional dummy variable. The economic and political factors which influence on dependent variable are represented by the control variables of the regression analysis. Due to the lack of appropriate data for all ASEAN countries for the selected time period, six member countries were selected on estimating impacts of ASEAN free trade agreement on food security of the region. Data collected for selected six member countries are Indonesia, Lao PDR, Malaysia, Philippines, Thailand and Vietnam.

*Specifications of Empirical Model*
Empirical model to estimate the relationship between RTAs and food security is given in equation (1). According to the model, it is assumed that national food security is measured in terms of overall food availability of the economy. According to literature food availability is represented by per capita daily energy supply. The model presented in equation (1) consider number of control variables which are significant in determining national level food security in addition to regional trade agreement (RTAs) dummy variable. Per capita real income, agricultural land, political instability, amount of foreign reserves available and import food prices are some important variables which can affect the level of food security of a country. By taking in to account all these matters, equation (1) is constructed to estimate impact of RTAs on food security of member countries of RTAs.

$$LogPES_{it} = \alpha_0 + \alpha_1 RTA_{it} + \alpha_2 LogY_{it} + \alpha_3 LogAL_{it} + \alpha_4 LogMP_{it-1} + \alpha_5 LogFRM_{it-1} + \alpha_6 PS_{it} + \varepsilon_i + v_t + \mu_{it} \quad ---(1)$$

Where *PES* is Per capita daily dietary energy supply, *RTA* is regional trade agreement dummy variable which equals one when a nation joined to RTA and zero otherwise, A*L* is agricultural land area, *MP* is the price of imported food, *FRM* is foreign reserves in months of imports, *Y* is per capita real GDP, *PS* is political stability, $\varepsilon$ is country-specific, time-invariant fixed effect, $v$ is period-specific, individual-invariant fixed effect, *m* is order of lag, $\mu$ is stochastic error term. Subscript *i* and *t* denote country and time respectively.

## 3. RESULTS

As mentioned in the previous section, the study defined food security as the food availability which is measured by the per capital daily dietary energy supply (PES). In determining the impacts of ASEAN free trade agreement on food security of member countries, behavior of PES before, during and after the implementation of discriminatory trade policies can be useful to study and it can give a fundamental inspiration. In examining the behavior of PES the study period is partitioned accordingly. The first period before the 1993 represent before discriminatory liberalization episode of ASEAN. All present members of ASEAN have signed on AFTA agreement by 2000 and therefore period after 2000 can be considered as after liberalization episode. The Period from 1992 to 1999 is regarded as the transition period since some of the present member countries were getting the membership of ASEAN organization.

Figure 1 depicts the long term trend of average PES of ASEAN member countries and CJK which represents China, Japan and Korea Republic. Over the last three decades average PES of ASEAN members shows an upward trend with a structural change. From 1980 to the beginning of 1990s, average PES shows mostly a slight upward trend. However, from the beginning of 1990s to present, average PES has been increased moderately showing an upward trend. For example, in 1980 average PES was 2195 and this figure has increased to 2199 by1990 from 4 points. However, from 1990 to 2000 average PES has been increased by almost by 240 points reaching to 2440 in 2000. In the next ten year period, from 2000 to 2009 PES has been increased from 2440 to 2669 by almost 229 points.

Figure 1: Average PES of ASEAN and CJK Countries

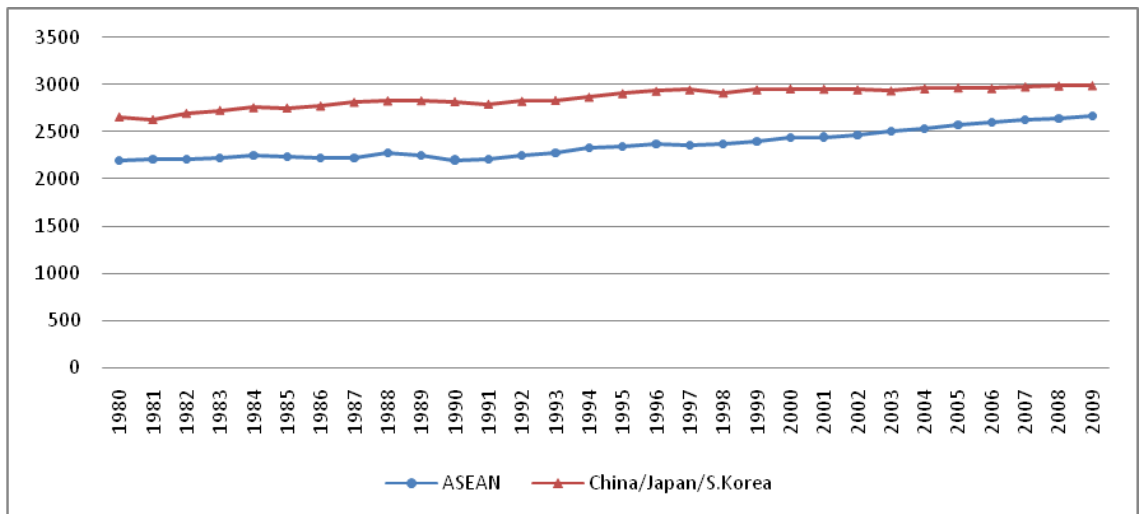

Source: Food and Agriculture Organization of the United Nations Statistical Database (FAOSTAT)

This increase of PES during the last decade is a significant change in the ASEAN region. Also, another important characteristics shown in Figure 1 is that gap between two PES lines has being decreased over the years due to higher growth of PES of ASEAN members compared to CJK countries. Figure 2 shows average PES of ASEAN members for several time periods of last three decades. Comparatively, average PES level of first two time periods shows same level without any significant change. However, three time periods belong to last two decades does show a significant increase of PES. This significant increase of PES probable might be due the implementation of the discriminatory trade liberalization policies with the formation of AFTA.

Figure 2: Average PES of ASEAN Members

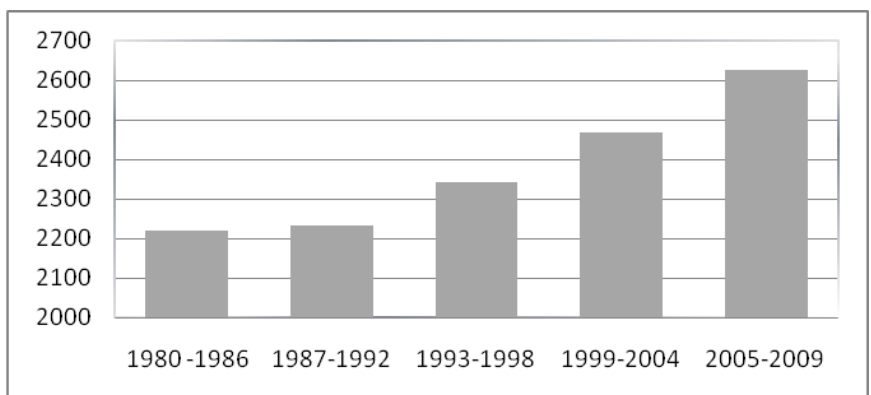

Source: Food and Agriculture Organization of the United Nations Statistical Database (FAOSTAT)

Figure 3 shows the average growth of PES of ASEAN members for selected periods. First two period of the figure, 1980-86 and 1987-92, represent before the implementation of AFTA agreement and last three periods mostly represent after implementation of AFTA trade agreement. PES growth rate during the first two selected period was only 0.2 percent. Last three periods comparatively shows higher growth rate of PES. For example, in 1993-1988, 1999-2004 and 2005-2009 periods, average PES of ASEAN ten member countries has been grown at 0.93 percent, 1.12 percent and 1.02 percent respectively.

Figure 3: Average Growth Rate of PES of ASEAN Members

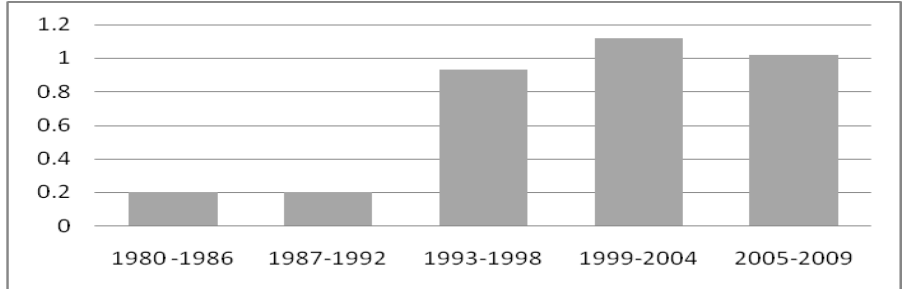

Source: Food and Agriculture Organization of the United Nations Statistical Database (FAOSTAT)

**Regression Results**

In estimating RTA impact on food security, multiple-regression is employed for panel data. The study used panel data for 6 six ASEAN member countries from 1980 to 2009. The model is estimated by using the STATA statistical software and summarizes results in Table 1 and 2.

Regression results for fixed effect model and random effect model are shown in column I and II of Table 1 respectively. All variables employed in the model except political stability and RTA dummy variable are expressed in logarithm. The independent variables employed in both methods are jointly significant. Fixed effect model shows approximately 30 percent of the variation in the dependent variable and random effect model explains approximately 50 percent variation in the dependent variable by explanatory variables.

Table 3: Basic Panel Regression Results
Dependent Variable: LogPES

| Independent Variable | I: Fixed Effect | | II: Random Effect | |
|---|---|---|---|---|
| | Coefficient | t-Statistics | Coefficient | t-Statistics |
| LogY | -0.0001174 | 0.04 | 0.0109587 | 2.73*** |
| LogAL | -0.0212943 | 0.34 | 0.0074527 | 0.96 |
| LogFRMt-1 | 0.0120204 | 2.90*** | 0.0146943 | 2.98*** |
| LogMPt-1 | -0.0183508 | 0.96 | -0.0365142 | 1.32 |
| PS | -0.0002433 | 0.19 | 0.0029606 | 2.34** |
| RTA | 0.1079062 | 8.29*** | 0.0825208 | 5.19*** |
| | | | | |
| N | 174 | | 174 | |
| Adj. R2 | 0.2953 | | 0.4991 | |
| | F = 44.61*** | | Wald Chi$^2$ =166.40*** | |

Note: The t-statistics are absolute values of t-ratios. Single, double and triple asterisks denote significance at the 10 %, 5% and 1 % level, respectively.

In fixed effect model, the signs of estimated coefficients of FRM and RTA dummy variable are as expected and significant at any standard significant level. However, under fixed effect model, coefficients of other explanatory variables are not statistically significant and their signs are not as expected except FRM and MP. However, the sign of the coefficient of MP is as expected though that coefficient is not significant. Colum II summarizes the results of random effect model. With compared to the results of fixed effect model random effect model shows further improvement as far as the statistical significance and signs of the variables are concerned. Coefficients of Y, FRM and RTA are statistically significant at 1 percent level while coefficient of PS is significant at 5 percent level. These three variables significantly influence on the dependent variable, PES. Signs of Each coefficient of independent variables under random effect model are as expected. That is all independent variables except MP are positively related with the dependent variable.

However, an examination of residuals obtained from the regressions under Fixed and Random effects shows the presence of serial correlation. With compared to time-series data, major advantage of panel data analysis is the increase of numbers of observations and the degree of freedom. However, panel data analysis may create some possibility of suffering in cross-sectional heteroskedasticity and can influence on the efficiency of the estimators. The results of the Breusch-Pagan test for heteroskedasticity indicate that its presence cannot be ruled out at the conventional level of significance. As a remedy for autocorrelation and heteroskedasticity, the model was re-estimated using feasible generalized least square (FGSL) and results are summarized in Table 2.

Table 4: Modified Panel Regression Results
Dependent Variable: LogPES

| **Independent Variable** | **I: FGLS** | | **II: PCSE** | |
|---|---|---|---|---|
| | **Coefficient** | **t-Statistics** | **Coefficient** | **t-Statistics** |
| LogY | 0.0109587 | 2.79*** | 0.0109587 | 2.92*** |
| LogAL | 0.0074527 | 0.98 | 0.0074527 | 2.13** |
| LogFRMt-1 | 0.0146943 | 3.04*** | 0.0146943 | 4.05*** |
| LogMPt-1 | -0.0365142 | 1.35 | -0.0365142 | 1.47 |
| PS | 0.0029606 | 2.39** | 0.0029606 | 3.07*** |
| RTA | 0.0825208 | 5.30*** | 0.0825208 | 6.45*** |
| | | | | |
| N | 174 | | 174 | |
| R-Squared | -- | | 0.4991 | |
| Overall Significance | Wald Chi$^2$ =173.37*** | | Wald Chi$^2$ =237.21*** | |

Note: 1.   the t-statistics are absolute values of t-ratios. Single, double and triple asterisks denote significance at the 10 %, 5% and 1 % level, respectively.
     2.    FGLS = Feasible generalized leased square, PCSE = panel-corrected standard error estimates, N = number of observations.

According to the regression results summarized in Table 2, it seems that correction for autocorrelation and heteroskedasticity has improved the statistical significance of the most of the independent variables. Results shows in column I for generalized least square estimates shows that signs of all coefficients are as expected and all independent variables except AL and MP are statistically significant. As a further improvement, column II shows the results of panel-corrected standard error estimates which are computed based on Prais -Winsten transformation by using *xtpcse* stata command instead of *xtgls*. Prais - Winsten transformation (PCSE) has further improved the significance of regression coefficients and explains approximately 50 percent variation of dependent variable. Results in column II shows that all explanatory variables except MP significantly influence on food security. The sign of the coefficient of variable MP, which represents imported food unit prices, is as expected though that coefficient is not statistically significant. Most importantly, the RTA dummy variable, which represents membership of ASEAN free trade agreement over the time significantly influence on food security of the member states.

## 4. CONCLUSION

The objective of the paper was to investigate whether ASEAN free trade agreement has influenced on food security of its member countries. With that objective, research was carried out to capture the relationship between food security and AFTA. In assessing the association between food security and AFTA free trade agreement, multiple regression analysis has been employed. In this study the food security is defined as the food availability of economies which is measured by per capita daily dietary energy supply (PES). The dependent variable is regressed with several explanatory variables which represent socioeconomic, geographic and political factors. The study employed RTA dummy variable to capture the influence of AFTA on food security of its member countries. Per capita real GDP, agricultural land area, amount of foreign reserves, imported food prices and political stability are the key control variables which was included in the regression model.

The major finding of the study is that AFTA has influenced positively in reducing food insecurity of its members. After the formation of AFTA, level of per capita daily dietary energy supply of the member countries shows moderate increase over the time. In addition to the trade effect on food security, sign of the political stability coefficient indicate that people's level of food security is influenced by the political stability of countries. Further study found that per capita income, agricultural land and level of foreign reserves have significantly influenced on food security of people of ASEAN nations. The literature of the study identified that imported food prices as a key determinant to influencing on food security. However, coefficient of imported food price of the study was not statistically significant. Nevertheless, the sign of the coefficient was as expected. Although the study does provide valuable information on RTA and food security, it does possess a limitation. The study has incorporated only food availability to measure food security of people and has not incorporated food accessibility and food stability, which are the other key pillars in determining food security of people. Therefore, future research needs to incorporate these three pillars on estimating food security.